\documentclass[10pt]{article}

\usepackage[utf8]{inputenc}
\usepackage[english]{babel}
\usepackage{epsfig}
\usepackage{amsmath}
\usepackage{amssymb}
\usepackage{color}
\usepackage{bbm}
\usepackage{hyperref}
\usepackage{physics}
\usepackage{graphicx}
\usepackage{xcolor,colortbl}
\usepackage{underscore}
\usepackage{cite}
\usepackage{listings}

\newcommand{\be}{\begin{equation}}
\newcommand{\ee}{\end{equation}}
\newcommand{\ba}{\begin{eqnarray}}
\newcommand{\ea}{\end{eqnarray}}
\newcommand{\bs}{\begin{subequations}}
\newcommand{\es}{\end{subequations}}
\newcommand{\no}{\nonumber\\}

\makeatletter
\def\hlinewd#1{%
	\noalign{\ifnum0=`}\fi\hrule \@height #1 %
	\futurelet\reserved@a\@xhline}
\makeatother
 
\lstdefinelanguage{Math14}
{morekeywords={},
sensitive=false,
morecomment=[l]{//},
morecomment=[s]{(*}{*)},
morestring=[b]",
}

\definecolor{codegreen}{rgb}{0,0.6,0}
\definecolor{codegray}{rgb}{0.95,0.95,0.95}
\definecolor{codepurple}{rgb}{0.0,0,0.8}
\definecolor{backcolour}{rgb}{0.95,0.75,0.92}
\definecolor{codecyan}{rgb}{0.29, 0.62, 0.73}
\definecolor{codemagenta}{rgb}{0.95,0.0,0.0}

\lstdefinestyle{mystyle}{
    backgroundcolor=\color{codegray},   
    commentstyle=\color{codecyan},
    keywordstyle=\color{codepurple},
    stringstyle=\color{gray!70!black},    
    basicstyle=\ttfamily\footnotesize,
    breakatwhitespace=false,         
    breaklines=true,                 
    captionpos=b,                    
    keepspaces=false,                 
    showspaces=false,                
    showstringspaces=false,
    showtabs=false,                  
    tabsize=2,
    lineskip=0.8ex
}
\graphicspath{ {figures/} }

\begin{document}

\title{\LARGE
  Conditions for boundedness from below
  of a $\Delta(54)$-symmetric three-Higgs-doublet model}

\author{
  Darius Jur\v{c}iukonis$^{(1)}$\thanks{E-mail:
    \tt darius.jurciukonis@tfai.vu.lt}
  \ and
  Lu\'\i s Lavoura$^{(2)}$\thanks{E-mail:
    \tt balio@cftp.tecnico.ulisboa.pt}
  \\*[3mm]
  $^{(1)}\!$
  \small Vilnius University, Institute of Theoretical Physics and Astronomy, \\
  \small Saul\.etekio~av.~3, Vilnius 10257, Lithuania
  \\*[2mm]
  $^{(2)}\!$
  \small Universidade de Lisboa, Instituto Superior T\'ecnico, CFTP, \\
  \small Av.~Rovisco~Pais~1, 1049-001~Lisboa, Portugal
}

\maketitle

\large

\begin{abstract}
  We investigate the orbit space of the scalar potential
  of a $\Delta(54)$-symmetric three-Higgs-doublet model.
  We find that,
  if the potential enjoys $CP$ invariance,
  then its three-dimensional orbit space is a polytope;
  if the potential has no $CP$ symmetry,
  then its four-dimensional orbit space has a boundary
  that is sometimes slightly concave,
  but seems never to be convex.
  Consequently,
  we conjecture necessary and sufficient conditions for the potential
  to be bounded from below;
  brute-force minimization of a large number of potentials
  affirms the accuracy of our conjecture.
  We list all possible charge-conserving and charge-breaking
  minima of the potential.
\end{abstract}

\newpage

\section{Introduction}
\label{sec:intro}

The Standard Model (SM) has gauge group $SU(2) \times U(1)$
and its scalar sector consists of just one $SU(2)$ doublet.
An obvious extension of the SM consists in enlarging its scalar sector.
Two-Higgs-doublet models have endured intense scrutiny~\cite{sher};
in this paper we deal on three-Higgs-doublet models (3HDMs).
Their scalar sector consists of three $SU(2)$ doublets $\Phi_k$ ($k = 1, 2, 3$),
all of them with the same nonzero $U(1)$ hypercharge.
We use an $SU(2) \times U(1)$ transformation to make
the upper component of $\Phi_1$ equal to zero,
and thereafter to make both the lower component
of $\Phi_1$ and the upper component of $\Phi_2$ non-negative real;
one then has
\bs
\label{peo}
\ba
\Phi_1 &=& \sqrt{N} \cos{\alpha} \left( \begin{array}{c} 0 \\
  1 \end{array} \right),
\\
\Phi_2 &=& \sqrt{N} \sin{\alpha} \cos{\beta} \left( \begin{array}{c}
  \sin{\gamma} \\ e^{i \psi} \cos{\gamma}
\end{array} \right),
\\
\Phi_3 &=& \sqrt{N} \sin{\alpha} \sin{\beta} \left( \begin{array}{c}
  e^{i \nu} \sin{\delta} \\ e^{i \xi} \cos{\delta}
\end{array} \right),
\ea
\es
where
\be
\label{N}
N := \sum_{k=1}^3 \Phi_k^\dagger \Phi_k
\ee
and $\alpha$,
$\beta$,
$\gamma$,
and $\delta$ are angles of the first quadrant.
We define $A_{kl} := \Phi_k^\dagger \Phi_l$.
Evidently,
$A_{lk} = A_{kl}^\ast$.
Note that $N \ge 0$ and that $A_{kl} A_{lk} \ge 0\ \forall l \neq k$.
The scalar potential of a 3HDM is $V = V_2 + V_4$,
where $V_2 = \sum_{k,l=1}^3 \mu_{kl} A_{kl}$
and $V_4 = \sum_{k,l,m,n=1}^3 \lambda_{klmn} A_{kl} A_{mn}$.
This potential has many coefficients.
Therefore,
it is convenient to supplement 3HDMs with non-gauge
symmetries,
which reduce the number of
coefficients
in $V$.
An example of one such symmetry is $\Delta (54)$,
which is generated
by the transformations\footnote{The transformations~\eqref{tra}
generate the group $\Delta (27)$,
which is smaller than $\Delta (54)$.
However,
since $V$ is renormalizable,
\textit{i.e.}\ it has no terms with more than four doublets,
a $\Delta (27)$-invariant potential is automatically invariant
under the transformation $\Phi_2 \leftrightarrow \Phi_3$ too,
\textit{cf.}\ Eqs.~\eqref{QRP},
and therefore it is automatically $\Delta (54)$-invariant.}
\bs
\label{tra}
\ba
\label{5}
& & \Phi_1 \to \Phi_2 \to \Phi_3 \to \Phi_1;
\\
\label{6}
& & \Phi_2 \to \omega \Phi_2,\ \Phi_3 \to \omega^2 \Phi_3,
\ea
\es
where $\omega := \exp \left( 2 i \pi / 3 \right)$.\footnote{Notice that
we do not specify the transformation rules,
both under the gauge symmetry and under transformations~\eqref{tra},
of the fermions present in a specific 3HDM.
This is because
here
we just care about the scalar potential.}

Let us define
\bs
\label{QRP}
\ba
Q &:=& A_{11} A_{22}+ A_{11} A_{33} + A_{22} A_{33},
\\
R &:=& A_{12} A_{21} + A_{13} A_{31} + A_{23} A_{32},
\\
P &:=& A_{12} A_{13} + A_{21} A_{23} + A_{31} A_{32}.
\ea
\es
These three quantities,
and also $N$ in Eq.~\eqref{N},
are invariant under transformations~\eqref{tra}.
Both $Q$ and $R$ are real,
while $P$ is in general complex.
The most general
$\Delta(54)$-invariant
potential is given by
\bs
\ba
V_2 &=& \mu N, \label{V2}
\\
V_4 &=& \lambda_1 N^2 + \lambda_2 \left( R - Q \right)
+ 2 \lambda_3 \left( \Re P - R \right)
+ \lambda_4 Q
+ 2 \sqrt{3}\, \lambda_5 \Im P,
\label{evo}
\ea
\es
where $\lambda_1, \ldots, \lambda_5$
are real dimensionless coupling constants.
Defining
\be
x := \frac{R - Q}{N^2},\quad y := \frac{2 \left( \Re P - R \right)}{N^2},\quad
q := \frac{Q}{N^2},\quad t := \frac{2 \sqrt{3}\, \Im P}{N^2},
\label{xyqt}
\ee
and
\be
\hat \lambda := \lambda_1 + \lambda_2 x + \lambda_3 y + \lambda_4 q
+ \lambda_5 t,
\label{bar}
\ee
one has\footnote{Note that an $A_4$-symmetric 3HDM~\cite{vazao,buskin}
also has a potential of the form~\eqref{Vdo}---though
with a different $P$ and therefore
a different shape of
the orbit space.}
\be
V = \mu N + \hat \lambda N^2.
\label{Vdo}
\ee
We assume that $\mu$ is negative
so that the minimum of $V$ has nonzero values of the doublets.
If $\mu, \lambda_1, \ldots, \lambda_5,\ x,\ y,\ q$,
and $t$ are kept fixed and one only lets $N$ vary,
then
the minimum of the potential in Eq.~\eqref{Vdo} is given by
$- \mu^2 \left/ \left( 4 \hat \lambda \right) \right.$.
The task of finding the global minimum of $V$ thus reduces
to the task of finding the smallest possible value of $\hat \lambda$.
Note that $\lambda_1, \ldots, \lambda_5$ must be such that
$\hat \lambda$ is non-negative for all possible $x,\ y,\ q,$ and $t$;
else,
by infinitely increasing $N$ one would make $V$ in Eq.~\eqref{Vdo}
tend to $- \infty$ and $V$ would have no
global minimum.
There are thus two problems:
keeping
$\hat \lambda \ge 0$,
and finding the minimum $\hat \lambda$;
the first one is the problem of $V$ being `bounded from below' (BFB).

$\Delta (54)$ is a subgroup of a set of symmetry groups
that one may impose on $V$,
namely $\Delta (54) \times CP$,
which occurs when $\lambda_5 = 0$;
$\left[ U(1) \times U(1) \right] \rtimes S_3$,
which occurs when $\lambda_3 = \lambda_5 = 0$;
$\Sigma (36)$,
which occurs when $\lambda_4 = \lambda_5 = 0$;
and $U(3)$,
which occurs when $\lambda_3 = \lambda_4 = \lambda_5 = 0$.
So,
if one solves the above problems for a $\Delta (54)$-symmetric 3HDM,
one automatically has their solution for $\Delta (54) \times CP$-invariant,
$\Sigma (36)$-invariant,
$\left[ U(1) \times U(1) \right] \rtimes S_3$-invariant,
and $U(3)$-invariant 3HDMs.

In order to solve our problems we must investigate the allowed orbit space,
\textit{i.e.}\ the set of quartets $\left( x, y, q, t \right)$
that may result from arbitrary doublets~\eqref{peo};
if the orbit space is either concave or a polytope,
then the minimum must be located at a vertex of the space~\cite{degee}.

There have been studies of the $\Delta (54)$-invariant scalar potential
in the past~\cite{nishi,ivo1,ivo2,ivo3,ivo4,ivo5,kalinowski,keus}.
Those studies affirmed that the minimum of $V$ may lie
at one of the four inequivalent orbit-space points
\bs
\label{p12}
\ba
V_1:= & &
\left( x, y, q, t \right) = \left( 0,\ 0,\ 0,\ 0 \right);
\\
V_2:= & & \left( x, y, q, t \right)
= \left( 0,\ 0,\ \frac{1}{3},\ 0 \right);
\\
V_6:= & & \left( x, y, q, t \right)
= \left( 0,\ -1,\ \frac{1}{3},\ 1 \right);
\\
V_8:= & & \left( x, y, q, t \right)
= \left( 0,\ -1,\ \frac{1}{3},\ - 1 \right).
\ea
\es
Those points are realized by the following sets of doublets~\cite{gerard}:
\bs
\ba
& & \Phi_1 = \left( \begin{array}{c} 0 \\ 1 \end{array} \right),\
\Phi_2 = \left( \begin{array}{c} 0 \\ 0 \end{array} \right),\
\Phi_3 = \left( \begin{array}{c} 0 \\ 0 \end{array} \right)\
\mbox{for}\ V_1;
\\
& & \Phi_1 = \left( \begin{array}{c} 0 \\ 1 \end{array} \right),\
\Phi_2 = \left( \begin{array}{c} 0 \\ 1 \end{array} \right),
\Phi_3 = \left( \begin{array}{c} 0 \\ 1 \end{array} \right)
\mbox{for}\ V_2;
\\
& & \Phi_1 = \left( \begin{array}{c} 0 \\ 1 \end{array} \right),\
\Phi_2 = \left( \begin{array}{c} 0 \\ 1 \end{array} \right),\
\Phi_3 = \left( \begin{array}{c} 0 \\ \omega \end{array} \right)
\mbox{for}\ V_6;
\\
& & \Phi_1 = \left( \begin{array}{c} 0 \\ 1 \end{array} \right),\
\Phi_2 = \left( \begin{array}{c} 0 \\ 1 \end{array} \right),\
\Phi_3 = \left( \begin{array}{c} 0 \\ \omega^2 \end{array} \right)
\mbox{for}\ V_8.
\ea
\es
Furthermore,
points $V_{1,2,6,8}$ are~\cite{nishi}
the only possible minima of the $\Delta (54)$-invariant potential that
conserve the $U(1)$ of electromagnetism,
\textit{i.e.}\ minima of $V$ where one may sets
only the lower components of the $\Phi_k$ nonzero.\footnote{This means that
the $\Phi_k$ mimic singlets of $SU(2)$,
\textit{i.e.}\ that $A_{kl} A_{lk} = A_{kk} A_{ll}\ \forall k \neq l$,
\textit{i.e.}\ that $x = 0$.}
Three questions were left unanswered by all previous works:
\begin{enumerate}
\item What are the conditions on $\lambda_1, \ldots, \lambda_5$ that guarantee
  that $V_4$ is BFB?
\item What are the possible charge-breaking minima of $V$?
  \item What are the conditions on $\lambda_1, \ldots, \lambda_5$ that guarantee
  that a specific minimum of $V$ is the global minimum?
\end{enumerate}
Here
we strive to answer questions~1--3;
they have already been answered before
for the cases
of a $\left[ U(1) \times U(1) \right] \rtimes S_3$-symmetric 3HDM~\cite{faro}
and of a $\Sigma(36)$-symmetric 3HDM~\cite{yang},
but neither for the case of a general $\Delta(54)$-symmetric 3HDM
nor even for the general $CP$-invariant version.

The outline of this letter is the following.
In section~\ref{sec:poly} we
describe the polytope
which is the convex hull\footnote{The importance
of finding the convex hulls of orbit spaces
for deriving the BFB conditions has been emphasized in Ref.~\cite{raidal}.}
of the $\Delta(54)$-invariant scalar potential.
In section~\ref{sec:nonCP} we answer the three questions
for a $\Delta(54)$-invariant 3HDM;
we do the same in section~\ref{sec:CP}
for a $\Delta(54) \times CP$-invariant 3HDM.
%A brief section~\ref{sec:summary} concludes this letter.

\section{The polytope}
\label{sec:poly}

\begin{itemize}
\item Let
  \be
  \Phi_k = \left( \begin{array}{c} a \\ b \end{array} \right),
  \qquad
  \Phi_l = \left( \begin{array}{c} c \\ d \end{array} \right),
  \ee
  where $a$,
  $b$,
  $c$,
  and $d$ are complex numbers.
  Then,
  \ba
  A_{kk} A_{ll} - A_{kl} A_{lk} &=&
  \left( \left| a \right|^2 + \left| b \right|^2 \right)
  \left( \left| c \right|^2 + \left| d \right|^2 \right)
  - \left| a^\ast c + b^\ast d \right|^2
  \no &=& \left| a d - b c \right|^2,
  \ea
  which is non-negative.
  Therefore $Q \ge R$,
  hence $x \le 0$.
\item Since $A_{kl} A_{lk} \ge 0$,
  $R \ge 0$.
  Hence,
  $x + q \ge 0$.
\item We note that
  \be
  \left| A_{12} + A_{23} + A_{31} \right|^2 = R + 2 \Re P \ge 0.
  \ee
  Hence,
  $3 \left( x + q \right) + y \ge 0$.
\item We also note that
  \bs
  \ba
  \left| A_{12} + \omega A_{23} + \omega^2 A_{31} \right|^2
  &=& R - \Re P - \sqrt{3} \Im P,
  \\
  \left| A_{12} + \omega^2 A_{23} +
  \omega A_{31}
  \right|^2
  &=& R - \Re P + \sqrt{3} \Im P
  \ea
  \es
  are both non-negative.
  Therefore,
  $y \le t \le - y$.
\item The parameterization~\eqref{peo} of the doublets gives
  \be
  Q = N^2 \sin^2{\alpha} \left( \cos^2{\alpha}
  + \sin^2{\alpha} \cos^2{\beta} \sin^2{\beta} \right).
  \ee
  The maximum possible value of this quantity is $N^2/3$,
  which is achieved when $A_{11} = A_{22} = A_{33}$.
  Hence,
  $q \le 1/3$.
\item Defining $C_\theta = \cos^2{\theta}$ and $S_\theta = \sin^2{\theta}$
  for $\theta \in \left\{ \alpha, \beta, \gamma, \delta \right\}$,
  one has
  \ba
  Q - R &=& N^2 S_\alpha \left\{
  C_\alpha \left( C_\beta S_\gamma + S_\beta S_\delta \right)
  + S_\alpha C_\beta S_\beta \left[
    \vphantom{\sqrt{C_\gamma S_\gamma C_\delta S_\delta}}
    C_\gamma S_\delta
    \right. \right.\no & & \left. \left.
    + S_\gamma C_\delta
    - 2 \sqrt{C_\gamma S_\gamma C_\delta S_\delta}
    \cos \left( \xi - \psi - \nu \right) \right] \right\}.
  \label{uyv}
  \ea
  Maximizing the quantity in the right-hand side of Eq.~\eqref{uyv}
  as a function of $\alpha$,
  $\beta$,
  $\gamma$,
  $\delta$,
  and $\xi - \psi - \nu$,
  one finds that the maximum value of $Q - R$ is $N^2/4$,
  which is reached for a variety of configurations of the doublets.
  Thus,
  $x \ge - 1/4$.
\end{itemize}

So,
in the four-dimensional orbit space $\left( x, y, q, t \right)$,
the allowed region satisfies
\be
\label{ui}
- \frac{1}{4} \le x \le 0,\quad
3 x + y + 3 q \ge 0,\quad
q \le \frac{1}{3},\quad
y \le t \le - y.
\ee
Since inequalities~\eqref{ui} are linear in $x$,
$y$,
$q$,
and $t$,
they define a polytope in orbit space.
That polytope is given in detail
in \href{https://github.com/jurciukonis/4D-polytope}{https://github.com/jurciukonis/4D-polytope}; here it suffices to state that it has eight vertices:
those in Eqs.~\eqref{p12} plus
\bs
\label{p34}
\ba
V_3 := & &
\left( x, y, q, t \right) = \left( - \frac{1}{4},\ 0,\ \frac{1}{3},\ 0 \right);
\\
V_4 := & & \left( x, y, q, t \right)
= \left( - \frac{1}{4},\ 0,\ \frac{1}{4},\ 0 \right);
\\
V_5 := & & \left( x, y, q, t \right)
= \left( - \frac{1}{4},\ - \frac{1}{4},\ \frac{1}{3},\ \frac{1}{4}
\right);
\\
V_7 := & & \left( x, y, q, t \right)
= \left( - \frac{1}{4},\ - \frac{1}{4},\ \frac{1}{3},\ - \frac{1}{4}
\right).
\ea
\es
The vertices~\eqref{p34} have $x \neq 0$
and therefore they break electromagnetic charge,
\textit{i.e.}\ they can only be realized
if not all three $SU(2)$ doublets have only one nonzero component,
which moreover is the same for all three of them.
Indeed,
one reaches $V_{4, 3, 5,7}$ through
\bs
\label{vac4}
\ba
& & \Phi_1 = \left( \begin{array}{c} 0 \\ 1 \end{array} \right), \
\Phi_2 = \left( \begin{array}{c} 1 \\ 0 \end{array} \right), \
\Phi_3 = \left( \begin{array}{c} 0 \\ 0 \end{array} \right)
\ \mbox{for}\ V_4;
\\
& & \Phi_1 = \left( \begin{array}{c} 0 \\ 1 \end{array} \right), \
\Phi_2 = \textstyle{\frac{1}{2}}
\left( \begin{array}{c} \sqrt{3} \\ - 1 \end{array} \right), \
\Phi_3 = \textstyle{\frac{1}{2}}
\left( \begin{array}{c} - \sqrt{3} \\ - 1 \end{array} \right)
\ \mbox{for}\ V_3;
\\
& & \Phi_1 = \left( \begin{array}{c} 0 \\ 1 \end{array} \right), \
\Phi_2 = \textstyle{\frac{1}{2}}
\left( \begin{array}{c} \sqrt{3} \\ - \omega^2 \end{array} \right), \
\Phi_3 = \textstyle{\frac{1}{2}}
\left( \begin{array}{c} - \sqrt{3} \\ - \omega^2 \end{array} \right)
\ \mbox{for}\ V_5;
\\
& & \Phi_1 = \left( \begin{array}{c} 0 \\ 1 \end{array} \right), \
\Phi_2 = \textstyle{\frac{1}{2}}
\left( \begin{array}{c} \sqrt{3} \\ - \omega \end{array} \right), \
\Phi_3 = \textstyle{\frac{1}{2}}
\left( \begin{array}{c} - \sqrt{3} \\ - \omega \end{array} \right)
\ \mbox{for}\ V_7.
\ea
\es

Thus:
all points in orbit space are inside a polytope with eight vertices,
and those vertices belong to the orbit space themselves.

\section{Answer to the questions}
\label{sec:nonCP}

\begin{figure}[h!] 
 \centering
\begin{center}
\includegraphics[width=0.45\textwidth]{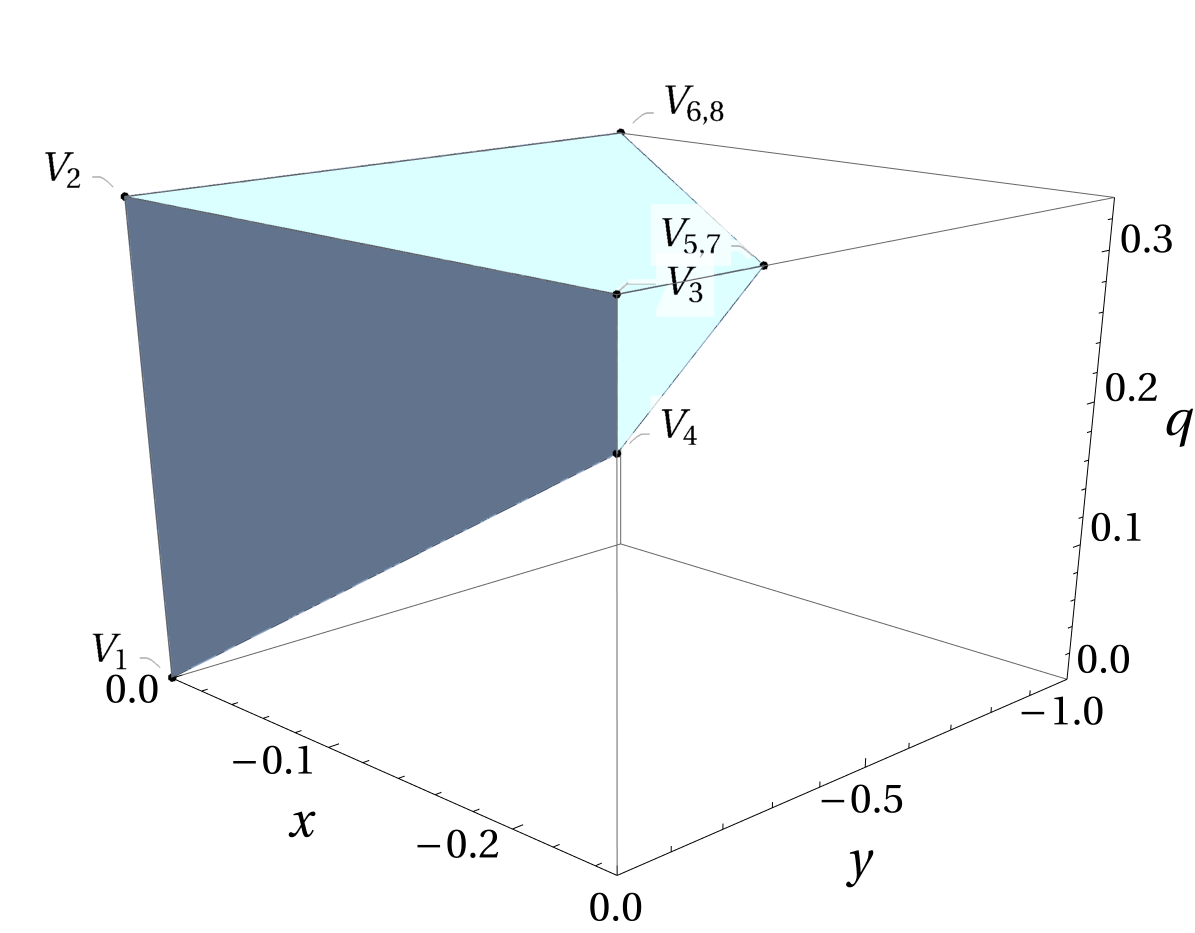}
\hspace{0.05\textwidth}
\includegraphics[width=0.4\textwidth]{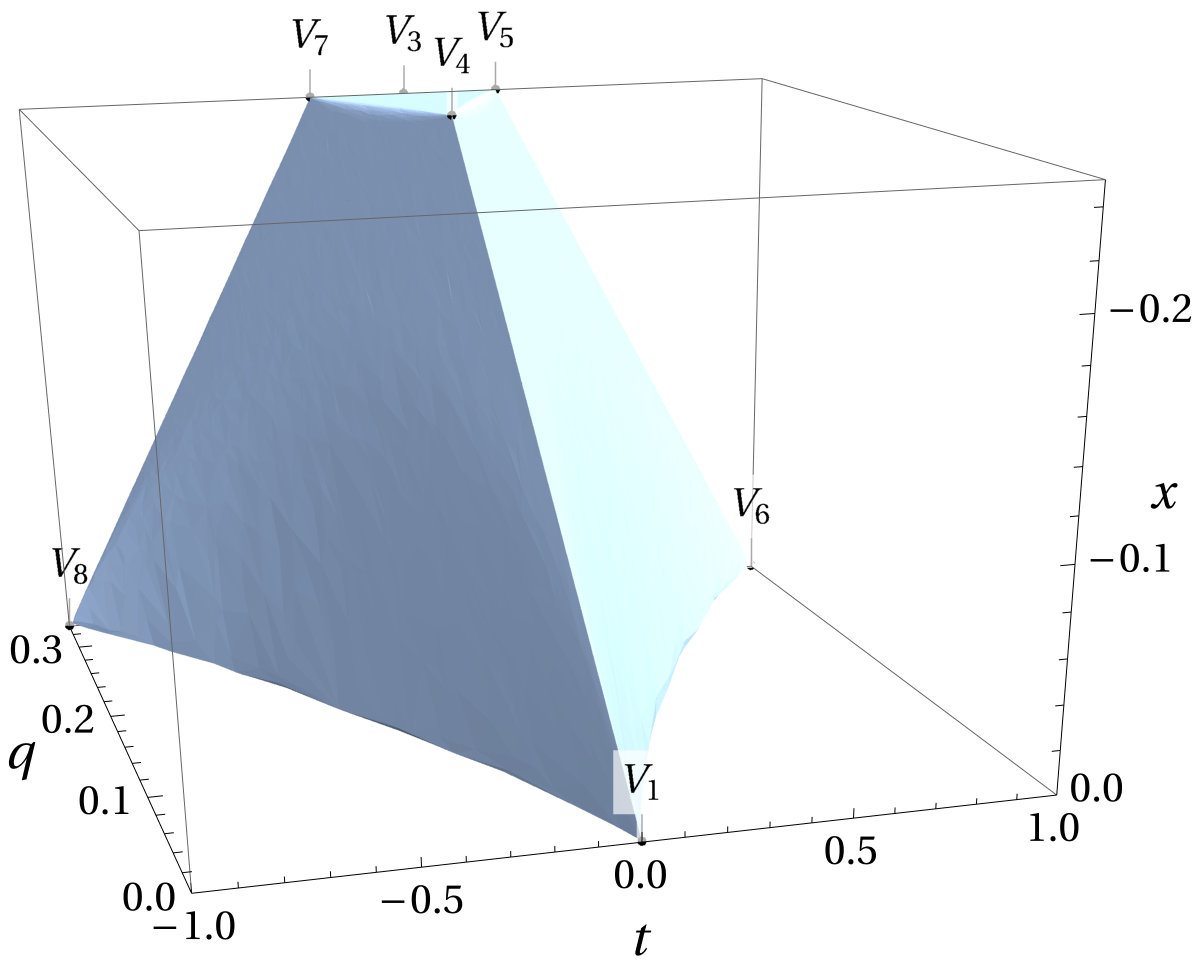}
\end{center}
 \caption{Left panel: a perspective of the projection
   of the orbit space onto the $\left( x, y, q \right)$ subspace.
   Right panel: same thing,
   but for a projection onto the $\left( x, q, t \right)$ subspace.
   The projections of the eight vertices $V_{1,\ldots,8}$
   of the four-dimensional polytope are displayed too
   ($V_2$ is hidden behind the volume in the right panel).}
  \label{fig:3projections}
\end{figure}

\begin{figure}[h!] 
 \centering
  %  \begin{tabular}{c}
%  \epsfig{file=varpi.eps,width=0.9\textwidth}
  \includegraphics[width=\textwidth]{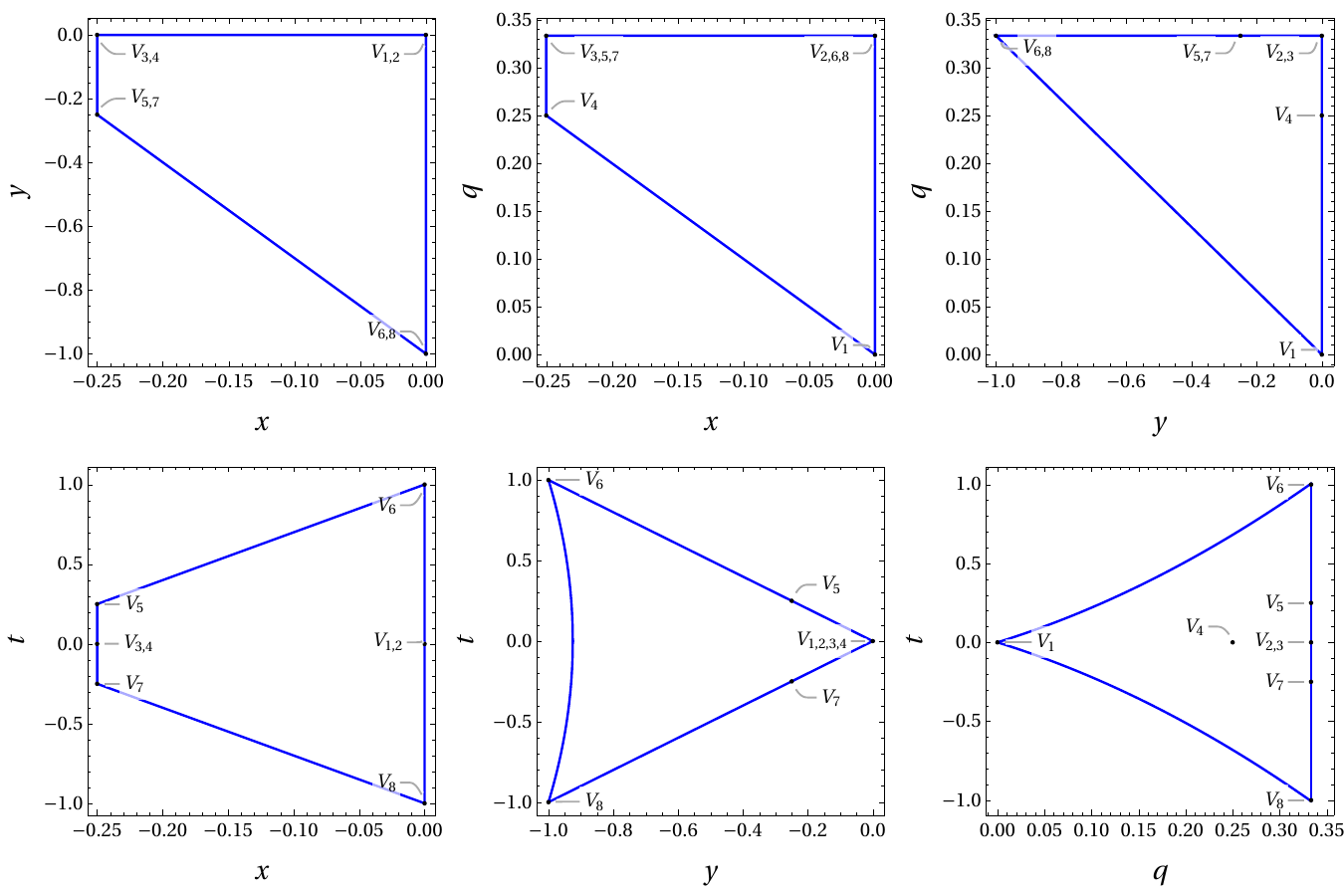}
%  \end{tabular}
  \caption{Six projections of
    the
    orbit space
   onto two-dimensional subspaces.
   The projections of the vertices
   of the four-dimensional polytope are indicated too.}
  \label{fig:2projections}
\end{figure}

One must investigate whether the orbit space coincides with the polytope,
or else
leaves
much of its volume un-realized.
In order to decide this we have randomly generated many
sets of values to the parameters $N$,
$\alpha$,
$\beta$,
$\gamma$,
$\delta$,
$\psi$,
$\nu$,
and $\xi$ of the doublets~\eqref{peo};
we have then used the resulting doublets to compute $x$,
$y$,
$q$,
and $t$.
We have found that the orbit-space points thus obtained almost fill
the entire polytope,
but not quite;
the allowed volume in orbit space appears to be slightly concave
in directions that involve the parameter $t$.
Note that in the preceding period we have used the phrase ``appears to be''
because we cannot visualize a four-dimensional space,
so we are not fully sure of our conclusion.
That conclusion is illustrated in Figs.~\ref{fig:3projections}
and~\ref{fig:2projections},
which display the projections of the four-dimensional orbit space
onto selected three- and two-dimensional,
respectively,
subspaces.

Since the convex hull of the orbit space is a polytope
and since the vertices of that polytope belong to the orbit space,
the possible minima of $V_4$ are at those vertices
and the questions posed at the end of the previous section
have the following answers:
\begin{enumerate}
\item Define
  \bs
  \ba
  \hat \lambda_1 &:=& \lambda_1,
  \\
  \hat \lambda_2 &:=& \lambda_1 + \frac{\lambda_4}{3},
  \\
  \hat \lambda_3 &:=& \lambda_1 - \frac{\lambda_2}{4} + \frac{\lambda_4}{3},
  \\
  \hat \lambda_4 &:=& \lambda_1 - \frac{\lambda_2}{4} + \frac{\lambda_4}{4},
  \\
  \hat \lambda_5 &:=& \lambda_1 - \frac{\lambda_2 + \lambda_3}{4}
  + \frac{\lambda_4}{3} + \frac{\lambda_5}{4},
  \\
  \hat \lambda_6 &:=& \lambda_1 - \lambda_3
  + \frac{\lambda_4}{3} + \lambda_5,
  \\
  \hat \lambda_7 &:=& \lambda_1 - \frac{\lambda_2 + \lambda_3}{4}
  + \frac{\lambda_4}{3} - \frac{\lambda_5}{4},
  \\
  \hat \lambda_8 &:=& \lambda_1 - \lambda_3
  + \frac{\lambda_4}{3} - \lambda_5.
  \ea
  \es
  The potential is BFB if and only if all the $\hat \lambda_i$
  ($i = 1, \ldots, 8$) are non-negative.
\item The possible charge-breaking minima of the potential
  are at the points~\eqref{p34} or,
  equivalently,
  at the orbits of the sets of doublets~\eqref{vac4}.
\item If one wants the global minimum of $V$ to be at
  any particular
  vertex $V_i$,
  then one must choose $\lambda_2, \ldots, \lambda_5$ such that
  $\hat \lambda_i < \hat \lambda_j\ \forall j \neq i$.
\end{enumerate}

In order to check the correctness of our conclusion~1,
we have randomly generated a data set comprising $10^6$ quartic potentials $V_4$
with couplings in the ranges $\lambda_1 \in [0, 10]$
and $\lambda_{2-5} \in [-10, 10]$.
We have minimized those $V_4$ by using brute force.
Without loss of generality,
we fixed
\be
\Phi_1 = \left( \begin{array}{c} 0 \\ 1 \end{array} \right),\
\Phi_2 = \left( \begin{array}{c} c \\ d + i e \end{array} \right),\
\Phi_3 = \left( \begin{array}{c} f + i g \\ h + i k \end{array} \right)
\ee
and we minimized $V_4$ with respect to the real numbers $c, d, e, f, g, h, k$.
The minimization procedure is described in Ref.~\cite{our};
we have used the {\tt \textsc{Mathematica}} function {\tt NMinimize}.
The minimization identified 561\,771 $V_4$
(out of the initial 1,000,000)
with
non-negative minimum.
Applying our analytical BFB conditions $\hat \lambda_i \ge 0\
\forall i = 1, \ldots, 8$,
we found exactly the same 561\,771 $V_4$.
This brute-force test demonstrates that the analytical BFB conditions
correctly identify all the BFB potentials.

\section{The $CP$-invariant case}
\label{sec:CP}

If the model has $\Delta (54) \times CP$ invariance,
then the last term in the right-hand side of Eq.~\eqref{evo} is absent.
This implies that the orbit space $\left( x, y, q \right)$
is three-dimensional instead of four-dimensional.
The polytope with eight vertices $V_1, \ldots, V_8$
becomes a polytope with just six vertices $\tilde V_1, \ldots, \tilde V_6$
because,
when one eliminates the parameter $t$,
the vertices $V_5$ and $V_7$ merge,
and the same happens to the vertices $V_6$ and $V_8$.
Thus,
\bs
\label{p56}
\ba
\tilde V_1: & & \left( x, y, q \right) = \left( 0, 0, 0 \right);
\label{p561}
\\
\tilde V_2: & & \left( x, y, q \right) = \left( 0, 0, \frac{1}{3} \right);
\label{p562}
\\
\tilde V_3: & & \left( x, y, q \right) =
\left( - \frac{1}{4}, 0, \frac{1}{3} \right);
\label{p563}
\\
\tilde V_4: & & \left( x, y, q \right) =
\left( - \frac{1}{4}, 0, \frac{1}{4} \right);
\label{p564}
\\
\tilde V_5: & & \left( x, y, q \right) =
\left( - \frac{1}{4}, - \frac{1}{4}, \frac{1}{3} \right);
\label{p565}
\\
\tilde V_6: & & \left( x, y, q \right) =
\left( 0, - 1, \frac{1}{3} \right).
\label{p566}
\ea
\es
The situation is simpler,
not just because the orbit space is three-dimensional
instead of four-dimensional,
and because the polytope has six vertices instead of eight,
but also because the orbit space \emph{completely fills} the polytope,
\textit{i.e.}\ it has flat boundaries everywhere instead of it having
some concave boundaries;
this can be seen in the left panel of Fig.~\ref{fig:3projections}
and in the top panels of Fig.~\ref{fig:2projections}.

Corresponding to the vertices $\tilde V_1, \ldots, \tilde V_6$ we define
\bs
\ba
\tilde \lambda_i &:=& \hat \lambda_i\quad
\forall i \in \left\{ 1, 2, 3, 4 \right\},
\\
\tilde \lambda_5 &:=& \lambda_1 - \frac{\lambda_2 + \lambda_3}{4}
+ \frac{\lambda_4}{3},
\\
\tilde \lambda_6 &:=& \lambda_1 - \lambda_3 + \frac{\lambda_4}{3}.
\ea
\es
Then:
\begin{enumerate}
\item The $\Delta (54) \times CP$-invariant scalar potential
  is bounded from below if and only if  all the $\tilde \lambda_i$
  ($i = 1, \ldots, 6$) are non-negative.
\item The possible charge-breaking minima
  of the $\Delta (54) \times CP$-invariant potential
  are at the
  points~\eqref{p563},
  \eqref{p564},
  and~\eqref{p565}.
\item If one wants the global minimum
  of the $\Delta (54) \times CP$-invariant potential
  to be at the vertex $\tilde V_i$,
  then one must choose $\lambda_2, \ldots, \lambda_5$ such that
  $\tilde \lambda_i < \tilde \lambda_j\ \forall j \neq i$.
\end{enumerate}
%

%\section{Summary}
%\label{sec:summary}

%The results of this work are summarized in the three points
%at the end of section~\ref{sec:nonCP}
%and in the three points at the end of section~\ref{sec:CP}.

%\vspace*{1mm}

\paragraph{Acknowledgements:} The work of D.J.\ received funding
from the Research Council of Lithuania (LMT)
under Contract No.~S-CERN-24-2;
part of the computations were performed
using the infrastructure of the Lithuanian Particle Physics Consortium
in the framework of agreement No.~VS-13 of Vilnius University with LMT.
The work of L.L.\ received funding from
the Portuguese Foundation for Science and Technology (FCT)
through projects UIDB/00777/2020 and UIDP/00777/2020,
and by the Recovery and Resilience Plan within the scope
of investment RE-C06-i06,
measure RE-C06-i06.m02,
project 2024.01362.CERN.

\end{document}